\documentclass[10pt, aps, pra, twocolumn, superscriptaddress, floatfix, showpacs, longbibliography]{revtex4-1}
\usepackage{graphicx}
\usepackage{dcolumn}
\usepackage{amsmath, amssymb}
\usepackage[dvipsnames]{xcolor}

\global\long\def\av#1{\left\langle #1 \right\rangle }

\global\long\def\hc{\hat{c}^{\phantom{\dag}}}

\global\long\def\hcd{\hat{c}^\dag}

\global\long\def\hn{\hat{n}}

\global\long\def\c{{\bm c}}

\global\long\def\part{\partial}
\global\long\def\Tr{{\rm Tr}}

\global\long\def\up{\uparrow}
\global\long\def\down{\downarrow}

\begin{document}
\preprint{APS/123-QED}
\title{Fluctuating Local Field Approach to Free Energy of 1D Molecules\\ With Strong Collective Electronic Correlations}

\author{Yana S. Lyakhova}
\affiliation{Russian Quantum Center, Skolkovo Innovation city, 121205 Moscow, Russia}
\affiliation{National Research Nuclear University MEPhI, 115409 Moscow, Russia}

\author{Evgeny A. Stepanov}
\affiliation{CPHT, CNRS, Ecole Polytechnique, Institut Polytechnique de Paris, F-91128 Palaiseau, France}

\author{Alexey N. Rubtsov}
\email{ar@rqc.ru}
\affiliation{Russian Quantum Center, Skolkovo Innovation city, 121205 Moscow, Russia}
\affiliation{Moscow State University, Physics Department, 119991 Moscow, Russia
}

\begin{abstract}
The impact of leading collective electronic fluctuations on a free energy of a prototype 1D model for molecular systems is considered within the recently developed Fluctuating Local Field (FLF) approach. 
The FLF method is a non-perturbative extension of a mean-field theory, where a self-consistent effective constant field is replaced by a fluctuating one.
Integrating the fluctuating field out numerically exactly allows one to account for collective electronic fluctuations mediated by this field without any assumptions on their magnitude, degree of non-linearity, etc.
Using a half-filled Hubbard ring as a benchmark system, we find that the FLF method 
noticeably improves a mean-field estimation for the free energy, in particular below the mean-field Ne\'el temperature.
We further demonstrate that the mean-field result can be even more improved introducing a multi-mode FLF scheme that additionally takes into account sub-leading fluctuations.
Possible applications for the thermodynamics of real molecules are also discussed. 
\end{abstract}
\maketitle

\section{Introduction}

Recent developments in the field of nano- and molecular electronics rely on finding effective low-dimensional systems that can be exploited to miniaturize electronic devices~\cite{doi:10.1021/acsanm.0c01386, Rueckes94, shulaker2013carbon, scheer2017molecular}.
The most prominent examples of such systems are quantum dots~\cite{PhysRevB.44.1646, goldhaber1998kondo, Cronenwett540, SCHMID1998182, RevModPhys.75.1, PhysRevLett.101.186804}, carbon nanotubes~\cite{tans1998room, javey2003ballistic, doi:10.1063/1.122477, RevModPhys.79.677}, grain boundaries and line defects in 2D systems like graphene~\cite{lahiri2010extended, huang2011grains, PhysRevLett.106.136806, Ebert_2014}, as well as single molecules, polymers and atomic chains~\cite{SHIRAKAWA19953, RevModPhys.73.713, doi:10.1021/jp011611w, khajetoorians2019creating}.
These nanoscale systems exhibit strong quantum effects and collective electronic fluctuations, which complicates their accurate theoretical description.
For instance, periodic one-dimensional (1D) systems possess collective modes such as solitons, polarons, and bipolarons~\cite{PhysRevLett.42.1698, RevModPhys.60.781}, as well as the Peierls instability, which appears already for an arbitrarily small electron-lattice interaction~\cite{PhysRevLett.55.308, PhysRevB.31.6633, PhysRevLett.62.2016, PhysRevLett.86.4572}.

A theoretical description of infinite interacting electronic systems is usually based on the notion of the free energy.
In particular, it allows to describe various phase transitions, such as the transition to magnetically ordered or superconducting states~\cite{landau1937theoryI, landau1937theoryII, ginzburg2009theory}. 
In the context of finite systems, an accurate estimation of the free energy of molecules is one of the central tasks for the quantum chemistry.
For instance, free energies of the reagents determine the equilibrium concentrations in chemical reactions.
A similar problem arises when calculating preferred molecular conformations, etc.   
While the free energy of small molecules can be obtained directly via the exact diagonalization or quantum Monte Carlo methods, the use of approximate calculation schemes for larger systems becomes unavoidable.
The most popular {\it state-of-the-art} approach used in material science and chemistry is the density functional theory (DFT) \cite{doi:10.1063/1.4869598}. 
DFT can be seen as a mean-field like method. By saying this we mean that within DFT a many-body problem of interacting electrons is mapped on an ensemble of non-interacting electrons living in an effective self-consistent potential.
The latter is adjusted to reproduce the density of the original interacting electronic problem. The accuracy of DFT is based on the construction of density functionals. Unfortunately, the exact form of the functionals is not known commonly, and usually the exchange-correlation part is the most challenging one. In practice it leads to approximations, such as the local density approximation (LDA) \cite{H-K_LDA}.
Despite the considerable success of the method in describing mean-field effects, this effective non-interacting approximation does not allow to capture collective many-body phenomena. 
For instance, DFT cannot capture formation of excitonic bands that appear as the result of electron-hole binding and can be revealed in optical spectra of molecules~\cite{Excitons}. 
A more sophisticated approximation for a many-body electronic problem relies on combining DFT with the dynamical mean-field theory (DMFT)~\cite{RevModPhys.68.13}. 
This allows for the exact numerical description of local many-body effects in the system including the local magnetic moment formation~\cite{stepanov2021spin} and the local renormalization of the spin-orbit coupling~\cite{PhysRevLett.116.106402, PhysRevB.97.085141, PhysRevLett.120.126401} and of the crystal-field splitting~\cite{10.1143/PTPS.160.233, PhysRevB.76.085127, PhysRevB.78.045115, PhysRevB.88.195116}.
However, this approach is not very suitable for effective 1D systems, because DMFT approximation becomes exact only in the limit of infinite spacial dimensions or connectivity of the lattice~\cite{PhysRevLett.62.324}.
In some cases, when the realistic 1D system can be approximated by the Hubbard model neglecting non-local electronic interactions, the exact solution for the effective 1D problem can be obtained exactly~\cite{PhysRevLett.20.1445, LIEB20031}.
At the same time, the Coulomb interaction in low-dimensional systems is usually long-ranged and weakly-screened, which imposes physical restrictions on this approximation.
In addition, collective electronic modes in molecular systems are essentially non-local. 
Typically, they involve a significant number of single-electron degrees of freedom and 
can be associated either with spin fluctuations, which is the case for molecular magnets, or with charge correlations seen, for example, in organic systems with $\pi$-bonds~\cite{PhysRevB.54.7965, doi:10.1080/15421400490481421, hermann2017nanoscale}.
DMFT neglects all non-local correlations by construction and thus does not allow to capture these collective electronic effects.

The existing set of theoretical tools for {\it ab initio} description of collective electronic fluctuations consists of calculating diagrammatic series for corresponding susceptibilities. 
The minimal approach yielding collective modes in weakly correlated systems is the random phase approximation (RPA)~\cite{QuantumSolidStatePhysics, LaundauFermiLiquid, TheoryLiquid}.
More advanced approximations handle collective degrees of freedom performing diagrammatic calculations on the basis of DMFT~\cite{RevModPhys.90.025003}.
Nevertheless, even these advanced methods cannot perform calculations well below the phase transition point predicted by DMFT~\cite{PhysRevX.11.011058}. 
A common problem of all diagrammatic schemes is an implicit assumption that collective fluctuations are small and linear, which allows to determine the leading (usually two-particle ladder-like) subset of diagrams.
This assumption works rather well above the mean-field estimation for the transition temperature. 
Lowering the temperature, the strength of collective fluctuations increases, and they become strongly non-linear. 
In particular this non-linearity can be explained by the fact that different collective modes start to interact with each other, which strongly affects the diagrammatic expansion~\cite{PhysRevB.94.035102, PhysRevB.96.035152, PhysRevB.102.195109, PhysRevB.103.245123}.

Recently, an alternative technique to handle collective modes dubbed ``Fluctuation Local Field'' (FLF) method~\cite{PhysRevE.97.052120, PhysRevB.102.224423} has been proposed. 
Within this approach, one or several collective modes can be accounted for numerically exactly.
An advantage of the proposed scheme is that no assumption about the magnitude and/or statistical properties of these fluctuations is made. 
Therefore, the FLF theory is expected to work well for systems, where the major part of strong fluctuations is comprised of several pronounced modes. In order to determine the leading modes in the system, one can use other (simpler) methods, such as the random phase approximation (RPA) \cite{TheoryLiquid,Platzman_Wolff}, the $GW$ approach \cite{PhysRev.139.A796,1998_Aryasetiawan,1999_Lars_GW}, or the fluctuation exchange (FLEX) method \cite{BICKERS1989206}. Once defined by means of these methods, leading collective modes can be successfully captured by the FLF approach.
In this regard, the FLF method looks promising for application to molecules, clusters, and nanostructures with developed collective modes.
In Ref.~\cite{PhysRevB.102.224423}, some of us have presented the FLF calculations of the magnetic susceptibility for small Hubbard plaquettes. 
Within these calculations the antiferromagnetic (AFM) mode, which represents the leading instability in the considered systems, has been accounted for by the FLF. 
It has been demonstrated that the FLF scheme describes the static response of the considered systems in a good agreement with the exact solution of the problem in the broad range of temperatures well below the limit of applicability of existing mean-field theories. 
However, other quantities of interest, e.g. thermodynamic potentials, have not been considered yet.
In this work we present the FLF calculations for a free energy of a prototypical molecular system that exhibits strong collective fluctuations, namely the half-filled periodized Hubbard chain. 
For simplicity, we restrict ourselves to a weakly-interacting case, so that the result of the Hartree-Fock (HF) method can be used for a comparison. 
We show that FLF method indeed improves the HF prediction for the free energy. 
Further, we demonstrate that increasing the number of fluctuating modes accounted for the FLF theory leads to a rapid improvement of the result.

\section{Variational principle for the Hubbard chain in the mean-field approximation}

In this work, we consider a Hubbard chain of $N$ lattice sites as a prototypical 1D molecular system.
The corresponding Hamiltonian reads:
\begin{align}
\hat{\cal H} = t\sum_{\langle ij \rangle,\sigma} \hcd_{i\sigma} \hc_{j\sigma}
+ U\sum_{j} \left(\hn_{j\up}-\frac12\right) \left(\hn_{j\down}-\frac12\right)
\label{H}
\end{align}
Here, $\hcd_{j\sigma}$ ($\hc_{j\sigma}$) is the creation (annihilation) operator for an electron at the lattice site ${j=0, \ldots N-1}$ with the spin projection ${\sigma=\{\uparrow, \downarrow\}}$. 
$t$ is the hopping amplitude between nearest-neighbor lattice sites $i$ and $j$ on which we impose the periodic boundary condition $\hc_{N}\equiv \hc_0$. 
$U$ is the on-site repulsive interaction between fermionic densities ${n_{j\sigma} = \hcd_{j\sigma}\hc_{j\sigma}}$ with opposite spin projections.
At half-filling the considered system exhibits strong AFM fluctuations, but the true ordering is never realized due to thermal and quantum zero-point fluctuations. 
In 1D these fluctuations are particularly important and have to be taken into account.
As anticipated in the Introduction, our goal is to obtain a free energy of the system, which can be expressed through the partition function ${\mathcal{Z} = \Tr{e^{-\beta \hat{\cal H}}}}$ as ${\mathcal{F} = -\beta^{-1} \ln \mathcal{Z}}$.
In the following we will also use the Lagrangian formalism, so one can write that ${\mathcal{Z} = \int D[c^{*}, c] e^{-{\cal S}[c^{*}, c]}}$, where ${{\cal S}[c^{*}, c]}$ is the corresponding action for the initial Hamiltonian~\eqref{H}.
In order to estimate the free energy of the system, we introduce a trial action ${\cal S}_{\rm tr}[c^{*}, c]$.
Then, ${\cal F}$ can be approximated by its first-order expansion in terms of the deviation of the initial problem form the trial action ${\cal S}-{\cal S}_{\rm tr}$:
\begin{align} 
\mathcal{F}\simeq-\beta^{-1} \ln \mathcal{Z}_{\rm tr}+\beta^{-1} \av{{\cal S}-{\cal S}_{\rm tr}}_{\rm tr}
\label{F1}
\end{align}
Here, the average $\av{\ldots}_{\rm tr}$ is taken with respect to the trial ${\cal Z}_{\rm tr}$ partition function. 
Should the trial action depend on some adjustable parameters, they can be chosen in such a way that their variation does not affect the free energy ($\delta {\cal F}=0$)~\eqref{F1}.
For a Hamiltonian trial system, this criterion represents the Gibbs-Bogoliubov-Feynman minimization principle~\cite{PhysRevE.66.066133}. 
Indeed, in this case it can be shown that the approximate free energy~\eqref{F1} reduces to ${\cal F} = \av{H}_{\rm tr}+\beta^{-1} \av{\ln \rho_{\rm tr}}_{\rm tr}$, where $\rho_{\rm tr}$ is the density matrix of the trial system. 
Then, the exact result ${\cal S}_{\rm tr}={\cal S}$ provides the lowest free energy limit for any approximate solution with ${\cal S}_{\rm tr}\neq{\cal S}$.
When the trial system is non-Hamiltonian, e.g. is non-local in time, the condition $\delta {\cal F}=0$ corresponds to the Peierls-Feynman-Bogoliubov variational principle~\cite{PhysRev.54.918, bogolyubov1958variation, feynman1972statistical}.
As we shall see below, in this case the free energy of the trial system is not necessarily higher than the exact one.

In this work we focus on the weakly interacting case, which allows to use the Hartree method as a parental one for the FLF approach. Following the mean-field idea, we first consider the simplest trial Hamiltonian where electrons interact only with an effective classical field $h$:
\begin{align}
\hat{\cal H}_{h} = t\sum_{\langle ij \rangle,\sigma} \hcd_{i\sigma} \hc_{j\sigma} - \sum_{j,l} h_{j}^l s^l_j
\label{HMF}
\end{align}
where ${s^l_j= \sum_{\sigma,\sigma'} \hcd_{j\sigma}\sigma^l_{\sigma\sigma'} \hc_{j\sigma'}}$ is the $l=\{x,y,z\}$ component of the spin density operator. 
Since leading collective electronic fluctuations in the initial problem~\eqref{H} are related to spin degrees of freedom, we consider only a site-dependent magnetic field $h^{l}_{j}$.
In this case, the minimization criterion $\delta {\cal F}=0$
leads to a well-known mean-field result:
\begin{align}
h_j^l=\frac{U}{2}\av{s^l_j}
\label{MFU}
\end{align}
that can be obtained self-consistently.
For the considered half-filled trial problem~\eqref{HMF}, this procedure predicts the SU(2) symmetry breaking below Ne\'el temperature ${T_{N} = \beta_N^{-1}}$.
The latter is associated with the formation of the AFM spin order and results in the following pattern ${\mathbf{h}_{j}=(-1)^j \,\mathbf{h}_{\rm AFM}}$ of the effective static AFM field $\mathbf{h}_{\rm AFM}$.
At higher temperatures the system remains paramagnetic, i.e. $\mathbf{h}=0$. 
As a matter of fact, this AFM ordering is an artifact of the approximation. As discussed above, a finite 1D system cannot possess a spontaneous symmetry breaking, which should be cured by a proper accounting for strong non-linear collective fluctuations.

\section{Fluctuating AFM local field}

Following the FLF idea presented in Refs.~\cite{PhysRevE.97.052120, PhysRevB.102.224423}, a mean-field artifact, namely the spontaneous symmetry breaking, can be avoided by considering fluctuations of the effective magnetic field.
To this aim we introduce a trial {\it ensemble} of mean-field problems~\eqref{HMF} subjected to different effective fields $h_j$ described by the following partition function:
\begin{align}
{\cal Z}_{\rm FLF} 
&= \sqrt{\det \frac{\beta N}{\Lambda}} \int
D[h_j] {\cal Z}_{h}  
\exp\Big\{-\frac{\beta N}{2}\sum_{i,j} h_i \Lambda^{-1}_{ij} h_j\Big\}
\label{Ztrial}
\end{align}
Here, ${\cal Z}_{h}$ is the partition function that corresponds to the mean-field problem~\eqref{HMF}, where now the effective magnetic field $h_j$ is considered as a variable. 
In general, this vector field $h_j$ may be different at different sites $j$, hence the integration is taken by ${D[h_j] = (2\pi)^{3N/2} \prod_{j} d^{3}h_j}$ 
with $N$ being the number of lattice sites. 
The magnitude of fluctuations, as well as the spatial pattern of the field $h_j$, is governed by the tensor $\Lambda_{ij}$, which will be determined later.

Since an exact accounting for many different fluctuating modes is numerically expensive, we consider only leading modes in actual calculations.
In this case it is convenient to rewrite the partition function in momentum-space representation performing a Fourier transform ${h_j=N^{-1} \sum_q h_q e^{i q r_j}}$. We also take into account the translational symmetry of the Hubbard ring, which allows one to write that ${\Lambda_{ij} = \Lambda_{i-j}}$.
It gives:
\begin{align}
{\cal Z}_{\rm FLF} = \int D[\tilde{h}_q] {\cal Z}_{h} \exp\Big\{-\frac12\sum_q  \tilde{h}_{q} \tilde{h}_{-q} \Big\}
\label{Ztrialq}
\end{align}
Here, ${D[h_q] = (2\pi)^{3N_q/2} \prod_{q} d^{3}\tilde{h}_q}$, where $N_q$ is the number of considered modes. 
Also, we rescaled the effective magnetic field as ${\tilde h_q = h_q \sqrt{\beta N \Lambda^{-1}_q}}$ to absorb the prefactor in Eq.~\eqref{Ztrial}.
Note that the partition function ${\cal Z}_{h}$ in this expression is written in terms of the original field $h_j$.
For a considered periodized chain~\eqref{H} leading magnetic fluctuations correspond to wave vectors ${q\approx\pi}$.
Keeping only a single AFM mode with ${q=\pi}$, one arrives at the simplest FLF realization considered in the previous paper~\cite{PhysRevB.102.224423}. In the present work, we extend the FLF approach to a multi-mode case and additionally consider two wave vectors ${q=\pi-\pi/N}$ and ${q=-\pi+\pi/N}$ that are nearest to the AFM mode.
This allows to take into account long-range spatial fluctuations of the AFM polarization, both in the magnitude and direction. 

It turns out that the resulting FLF problem~\eqref{Ztrialq} being written in terms of only fermionic variables is not local in time and thus is non-Hamiltonian.
Indeed, integrating out effective magnetic fields $h_j$ gives the following form for the partition function of the FLF problem
\begin{align}
{\cal Z}_{\rm FLF} = \int D[c^{*},c] e^{-{\cal S}_{\rm FLF}[c^{*},c]}
\label{Z_FLF2}
\end{align}
with an effective trial action~\cite{PhysRevB.102.224423}
\begin{align}
{\cal S}_{\rm FLF}[c^{*},c] = {\cal S}_0[c^{*},c] - \frac{1}{2\beta N} \sum_{q,l}  \int d\tau_1 d\tau_2\, \Lambda_{q} s_{\tau_1,q}^l s^l_{\tau_2,-q}
\label{SFLF}
\end{align}
where ${\cal S}_0[c^{*},c]$ is the non-interacting part of the initial action ${\cal S}[c^{*},c]$, and $s_q^l$ is a Fourier transform of $s^l_j$. 
We note that both, the initial and the trial FLF actions differ only in the interaction term. 
Therefore, to obtain the free energy~\eqref{F1} one only needs to calculate averages of these interaction terms with respect to the FLF partition function ${\cal Z}_{\rm FLF}$.
For calculating the average of the interaction part of the initial action is convenient to take the partition function in the form of Eq.~\eqref{Ztrialq}.
Then, the average over the FLF ensemble can be obtained as~\cite{PhysRevB.102.224423} 
\begin{align}
\av{\ldots}_{\rm FLF} = \int D[\tilde h_q] \av{\ldots}_h p_h   
\end{align}
where $p_h = \frac{{\cal Z}_h}{{\cal Z}_{FLF}} e^{-\frac12\sum_q  \tilde{h}_q \tilde{h}_{-q}}$.
The $\av{\ldots}_h$ stands for the average over ${\cal H}_h$, which is easy to calculate, because ${\cal H}_h$ is Gaussian in terms of fermionic variables. 
Then, the average of the Hubbard interaction term of the initial problem~\eqref{H} reads~\cite{PhysRevB.102.224423}
\begin{gather}
U \sum_j \av{\left(\hn_{j\up}-\frac12\right) \left(\hn_{j\down}-\frac12\right)}_{\rm FLF} = \notag\\
-\frac{UN}{4} \int D[\tilde{h}_q] \sum_{q}\av{{\bf s}_q}^2_{h}p_h = \notag\\ 
-\frac{U}{4 \beta^2 N}  \int D[\tilde{h}_q] \sum_q \left| \partial_{h_q} \ln {\cal Z}_h \right|^2  p_h
\end{gather}
The partial derivative that appears in this expression means $\partial_{h_q} \ln {\cal Z}_h = \frac{\partial \ln {\cal Z}_h}{\partial{}h_q}\Big|_{h_q=0}$.
The average of the interaction part of the FLF action~\eqref{SFLF} is convenient to take over the corresponding partition function~\eqref{Z_FLF2} as
\begin{gather}
- \frac{1}{2\beta N} \sum_{q,l}  \int d\tau_1 d\tau_2\, \Lambda_{q} \av{s_{\tau_1,q}^l s^l_{\tau_2,-q}}_{\rm FLF} = \notag\\
\sum_{q} \Lambda_{q} \, \partial_{\Lambda_{q}} \ln{\cal Z}_{\rm FLF}
\end{gather}
The estimation for the free energy~\eqref{F1} for the FLF trial action becomes
\begin{align}
{\cal F}\simeq 
&-\frac{1}{\beta} \left(1+\sum_q\Lambda_q  \partial_{\Lambda_q}\right) \ln {\cal Z}_{\rm FLF} \notag\\
&- \frac{U}{4 \beta^2 N}  \int D[\tilde{h}_q] \sum_q \left| \partial_{h_q} \ln {\cal Z}_h \right|^2  p_h 
\label{Fq}
\end{align}
Parameters $\Lambda_q$ that enter the derived expression will be defined below. 

It is worth noting that a straightforward justification of the Gibbs-Bogoliubov-Feynman minimization criterion does not apply in this case, because the FLF trial action~\eqref{SFLF} does not correspond to any Hamiltonian. 
Consequently,~\eqref{F1} cannot be rewritten as an average with some positive-defined density matrix, and $\mathcal{F}'$ does not appear to be lower-bounded by $\mathcal{F}$. 
However, our numerical analysis presented below shows that the function ${\cal F}(\Lambda)$ still has a minimum.
We argue that choosing $\Lambda$ at or near this minimum provides a good estimation for the free energy of the system. 

\section{Numerical procedure and results}

Let us turn to numerical results of the FLF approach. 
We perform calculations for periodic Hubbard chains of $N = 8$, 10, and 12 sites within single- and multi-mode FLF schemes.
Results for the free energy are compared with the mean-field (MF) estimation and the reference data obtained via the exact diagonalization (ED) method.
For the sake of applicability of the mean-field approximation we consider the regime of moderate electronic correlations and set $t=U=1$, so that the value of the on-site Coulomb potential is equal to the quarter of the bandwidth.

In the framework of the single-mode FLF scheme that involves only one adjustable parameter $\Lambda_{\pi} \equiv \lambda$ the partition function ${\cal Z}_{\rm FLF}$ can be obtained by the grid integration over the single $h_{\pi}$ variable.
In the multi-mode case that accounts for three classical vector fields the integration is taken over the 9-dimensional space, and the grid scheme is not applicable anymore. 
Instead, we randomly distribute about $10^7$ points in the $h_{q}$-space and estimate the integral by a sum over all these points with proper weighting factors.
The same value of ${\Lambda_q = \lambda}$ was taken for all three modes ${q=\pi, \pi\pm\pi/N}$, which is a reasonable choice for ${\pi/N \ll \pi}$.
The partial derivative with respect to $\Lambda_{q}$ in Eq.~\eqref{Fq} was calculated numerically in both, single- and multi-mode cases.

\begin{figure}[t!]
\includegraphics[width=0.95\linewidth]{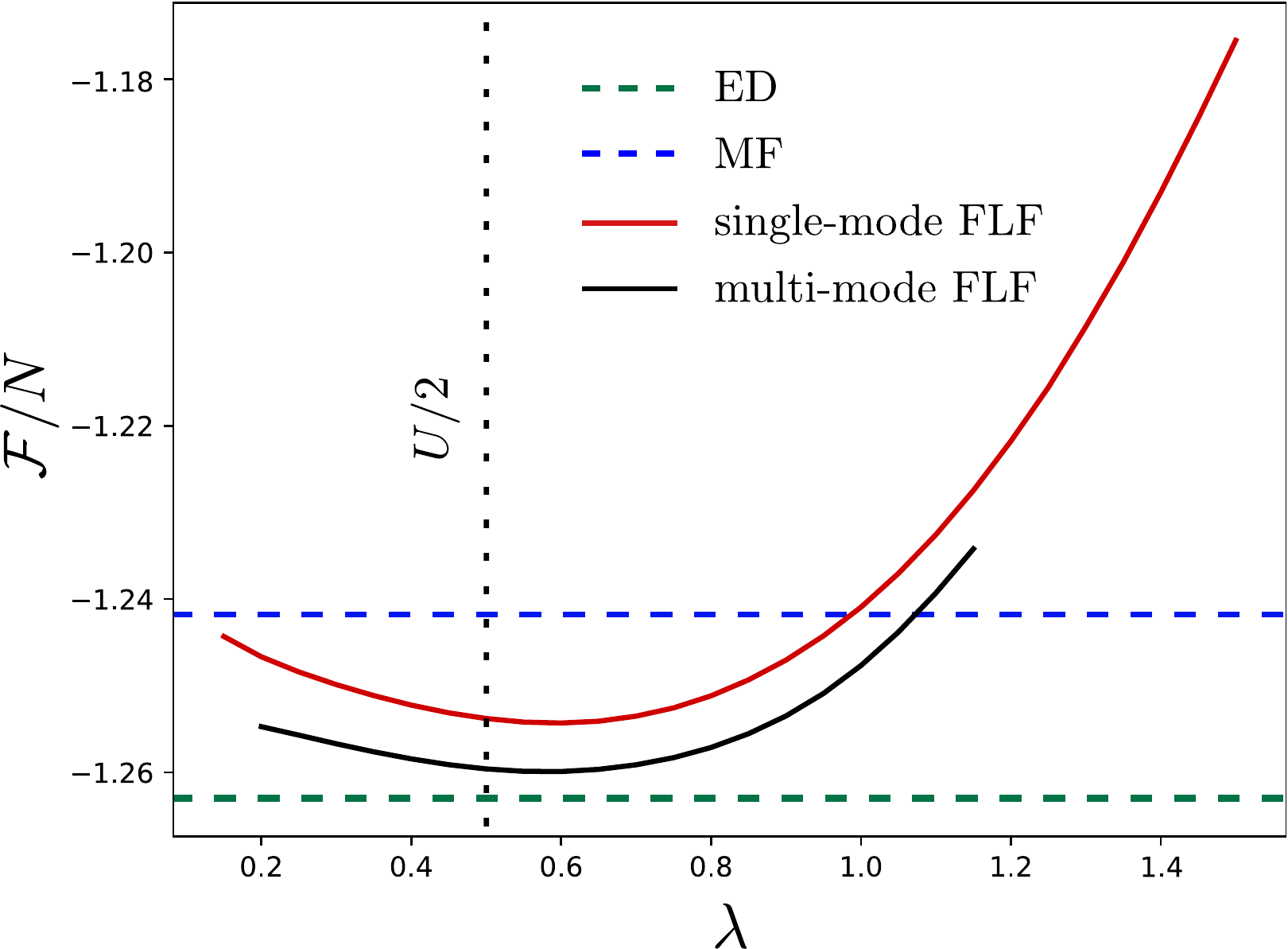}
\caption{\label{Fig1} Free energy ${\cal F}$ normalized by the number of sites ${N=8}$ obtained for ${t=U=1}$ and $\beta=10$ as a function of $\lambda$ parameter. In the single-mode case (red solid line) this parameter coincides with $\Lambda_{\pi}$. In the multi-mode case (black solid line) we take the same value ${\Lambda_{q} = \lambda}$ for all three considered modes ${q=\pi, \pi\pm\pi/N}$.
Vertical dashed black line indicates the saddle point estimation $\lambda=U/2$. Horizontal dashed blue and bold green lines correspond to MF and ED results, respectively.
}
\end{figure}

\begin{figure*}[t!]
\includegraphics[width=1\linewidth]{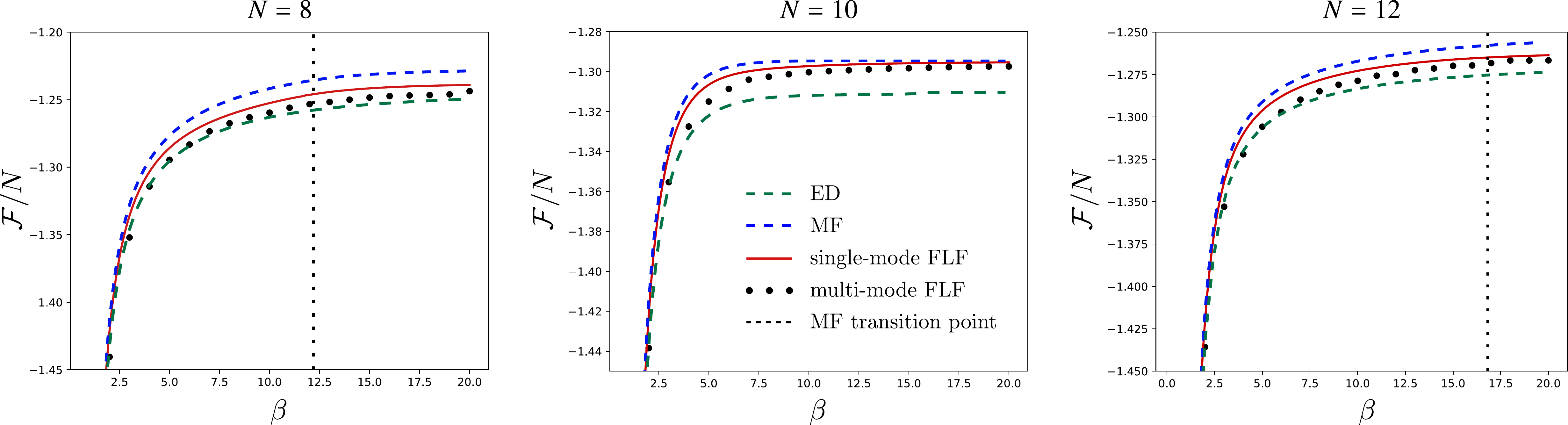}
\caption{\label{fig:F8-12} Free energy ${\cal F}$ as a function of the inverse temperature $\beta$ computed via ED (dotted green line), MF (dashed blue line), single-mode FLF (solid red line), and multi-mode FLF (black points) methods. Vertical dashed black lines indicate the transition temperature to the AFM ordered state predicted by MF approach.}
\end{figure*}

Let us first analyze the dependence of the free energy~\eqref{Fq} on $\Lambda$. 
Fig.~\ref{Fig1} shows the result for the free energy of the $N=8$ site system obtained within single- (solid red line) and multi-mode (solid black line) FLF schemes, as well as from the MF estimation (dashed blue line) and the ED method (dashed green line). 
It is worth noting that calculations have been performed for a relatively low temperature ($\beta=10$, MF transition point corresponds to $\beta^{\ast}\simeq12$), so that we expect that the system exhibits well-developed collective fluctuations.
We observe that the ${\cal F}(\Lambda)$ function has a minimum at $\Lambda_{min}\approx 1/2$, and the corresponding minimal value lies closer to the reference ED result than to the MF estimation. 
Although it has been pointed out that the Gibbs-Bogoliubov-Feynman criterion does not apply here, one can still find physical arguments to fix $\Lambda=\Lambda_{min}$. 
Generally speaking, it is expected that a ``good'' approximation that involves some free parameters should provide a result that weakly depends on these parameters. 
Remarkably, our theory satisfies this criterion, because the function ${\cal F}(\Lambda)$ shows a small change in its value within a relatively broad interval near $\Lambda_{min}$, in particular in the multi-mode case. 
If our theory was exact, the curve ${\cal F}(\Lambda)$ would be perfectly flat and $\Lambda$ would be a gauge field. Following this line of argumentation, the requirement $\Lambda=\Lambda_{min}$ ensures that ${\cal F}$ is at least locally independent of $\Lambda$. 
However, taking another value of $\Lambda$ that is close but not exactly equal to $\Lambda_{min}$ will yield nearly to the same result.

A particularly simple way to choosing $\Lambda$ can be found considering a saddle-point approximation for the integral over $h$ in Eq.~\eqref{Ztrial} as shown in Ref.~\cite{PhysRevB.102.224423}. 
Within this approximation the FLF ensemble of fields becomes replaced by a single constant field $h=\Lambda \av s$,
as it follows from the variation of the integrand in Eq.~\eqref{Ztrial}.
The requirement $\delta {\cal F}(\Lambda)=min$ in this case yields  
$
{h = \Lambda_{min}\av{s} = \frac{U}{2}\av{s}}
$,
which remarkably coincides with the MF solution~\eqref{MFU}.
The latter way of fixing $\Lambda$ is technically much simpler than finding an exact minimum and, as we observe from Fig.~\ref{Fig1}, leads to almost the same result for the free energy.

Now we turn to the temperature dependence of the free energy obtained within the FLF method. 
Fig.~\ref{fig:F8-12} shows corresponding results obtained for Hubbard chains with $N=8$, 10, and 12 sites using the single- and multi-mode FLF schemes with $\Lambda_{q}=U/2$. 
We find that accounting for collective fluctuations always improves the MF result, but the degree of the improvement at low temperatures is drastically better for systems with $N=8$ and $N=12$ sites than for the $N=10$ case. 
This fact has a simple physical explanation. 
For $N$ being a multiple of 4, a discrete momentum grid of the Brillouin Zone includes points $q=\pm\pi/2$. At these points the electronic spectral function of the considered model~\eqref{H} appears exactly at the Fermi level.
As a consequence, this results in a resonance in the density of states, which enhances the intensity of two-particle excitations that in our case correspond to AFM fluctuations.
On the contrary, in the system of $N=10$ sites the Fermi level falls between the $q=2 \pi/5$ and $q=3 \pi/5$ points of the momentum grid, which reduces the strength of the AFM fluctuations. 
This conclusion is also supported by the MF calculations that predict the transition to the AFM ordered state for $N=8$ and $N=12$ at certain temperatures marked by vertical lines in Fig.~\ref{fig:F8-12}, but does not reveal such a transition for the case of $N=10$ sites. 
AFM fluctuations that are captured by the FLF method are particularly strong below the transition point predicted by the MF theory.  
On the contrary, at high temperatures the effects of a discrete spectrum are smeared out, and we observe no essential difference between the results obtained for different number of lattice sites. 
We also note that for some temperatures the result obtained within the multi-mode FLF scheme lies below the reference ED data. 
This observation confirms our statement that the Peierls-Feynman-Bogoliubov variational principle is indeed not directly applicable to the FLF trial ensemble.

As can also be seen from Fig.~\ref{fig:F8-12}, taking into account multiple fluctuating modes results in a remarkable change of the FLF result, shifting it closer to the reference ED data. 
Introducing these additional modes allows one to account for different polarizations in different parts of the system. 
Our result suggest that this effect can be important even for relatively small molecules. 
In the current work we limited ourselves to the three leading $q$-modes.
Including a larger number of fluctuating modes is, in principle, possible for the cost of heavier numerical efforts. However, it should be noted that the choice $\Lambda_{q} \approx U/2$ is justified only for a few modes lying near the $q=\pi$ point. For other FLF schemes the problem of choosing a proper $\Lambda$ should be revisited.

\section{Conclusion and outlook}

To conclude, we demonstrated that the Fluctuating Local Field method significantly improves the mean-field results for the free energy of a 1D Hubbard chain that models molecular systems.
The multi-mode version of the FLF theory was introduced to simultaneously account for different fluctuations of the order parameter in different parts of the system.
We showed that including sub-leading modes has a noticeable effect on the free energy even for the case of small-sized systems. 
It can be expected that including more collective modes would lead to a further improvement of the results. 
This is especially important for larger systems, where the sub-leading fluctuating modes also become significantly important. 
At the same time, increasing the number of the modes tremendously rises computational costs. 
For this reason, considering a thermodynamic limit within the FLF approach can be seen as the future perspective.

In this work we restricted ourselves to a simple 1D model that
allow to compare the FLF result with the exact solution provided by the exact-diagonalization method. 
Density-functional theory (DFT) calculations for real molecular systems operate with multi-orbital Hamiltonians. 
In this case an exponential increase of the dimensionality of the Fock space makes the numerically exact solution of the problem impossible. 
On the contrary, the numerical costs of the single-mode FLF calculations is comparable to the mean-field ones regardless of the number of orbitals. 
This opens a perspective for a combined DFT+FLF treatment of
molecular systems with strong collective modes.

\section*{Acknowledgments}
The work of E.A.S. was supported by the European Union’s Horizon 2020 Research and Innovation programme under the Marie Sk\l{}odowska Curie grant agreement No.~839551 - $\text{2DMAGICS}$.

\bibliography{Ref}

\end{document}